\documentclass[aps,pra,twocolumn,showpacs,groupedaddress]{revtex4}  
\usepackage{graphicx}  
\usepackage{dcolumn}   
\usepackage{bm}        
\usepackage{amssymb}   
\usepackage{amsmath}
\usepackage{float}
\usepackage{ulem}
\usepackage{enumerate}
\usepackage{hyperref}
\begin{document}
	\title{Emerging chimera states under  non-identical  counter-rotating oscillators }
	\author{K. Sathiyadevi,$^{1}$,  V. K. Chandrasekar,$^{2}$,   and M. Lakshmanan$^{1}$}
	\address{$^1$Department of Nonlinear Dynamics, School of Physics, Bharathidasan University, Tiruchirappalli - 620 024, Tamil Nadu, India. \\$^1$Department of Physics, Centre for Nonlinear Science \& Engineering, School of Electrical \& Electronics Engineering, SASTRA Deemed University, Thanjavur -613 401, Tamil Nadu, India. }
	\begin{abstract} 
		\par    
		Frequency plays a crucial role in exhibiting various collective dynamics in the coexisting co- and counter-rotating systems. To illustrate the impact of counter-rotating frequencies, { we consider a network of non-identical and globally coupled Stuart-Landau oscillators with additional perturbation}. Primarily, we investigate the dynamical transitions in the absence of perturbation, demonstrating that the transition from desynchronized state to cluster oscillatory state occurs through an interesting partial synchronization state in the oscillatory regime. Followed by this, the system dynamics transits to amplitude death and oscillation death states. Importantly, we find that the observed dynamical states do not preserve the parity($P$) symmetry in the absence of perturbation. When the perturbation is increased one can note that the system dynamics exhibits a new kind of transition which corresponds to a change from incoherent mixed synchronization to coherent mixed synchronization through chimera state. 	In particular, incoherent mixed synchronization and coherent mixed synchronization states completely preserve the $P$-symmetry, whereas the chimera state  preserves the $P$-symmetry only partially. To demonstrate the occurrence of such partial symmetry breaking (chimera) state, we use basin stability analysis and discover that partial symmetry breaking  exists as a result of the coexistence of symmetry preserving and symmetry breaking behavior in the initial state space. Further, a measure of the strength of $P$-symmetry is established to quantify the $P$-symmetry in the observed dynamical states. Subsequently, the dynamical transitions are investigated in the parametric spaces. Finally, by increasing the network size, the robustness of the chimera state is also inspected and we find that the chimera state is robust even in networks of larger sizes. {We also show the generality of the above results in the related phase reduced model as well as in other coupled models such as the globally coupled van der Pol and R\"ossler oscillators.  }
		
	\end{abstract}
	\maketitle
	
	\section{Introduction}
	\par 
	\par 	Coupled nonlinear oscillators are effective tools for studying the collective behaviors of many complex dynamical systems in nature. The interaction of coupled elements, properties of an isolated system, and coupling topologies result in a variety of collective dynamical states such as desynchronization, synchronization, clusters, traveling waves, solitary states, and oscillation death states \cite{syn1,syn2, coh1,sat_vdp, adod}. Among all these states, the coexistence of spatially coherent and incoherent dynamical behavior known as ``chimera" has been investigated vigorously during the past couple of decades. It is a fascinating hybrid dynamical state having a strong resemblance with many real-world phenomena, including uni-hemispheric slow-wave sleep \cite{uni1,uni2,uni3}, bump states in neural networks \cite{pumb_neu}, power grid blackouts \cite{power1,power2}, social networks \cite{social}, etc.  The chimera state is also related to various brain diseases such as Parkinson’s disease, epileptic seizures, Alzheimer’s disease, schizophrenia, and brain tumors \cite{chim}. Originally, the coexistence of coherent and incoherent oscillatory dynamics has been reported in a network of nonlocally coupled oscillators by Kuramoto \cite{chim1}. Later, it was named as a chimera by Abrams and Strogatz \cite{chim2}. Eventually, the chimera state has also been realized in locally and globally coupled oscillators in various numerical models \cite{chim_rev1,ac_local,chim_global,prema_glob}. Also, a variety of chimera states including breathing chimera \cite{bc}, amplitude chimera \cite{sat_ac}, frequency chimera \cite{prema_glob}, traveling chimera \cite{tc1,tc2}, and imperfect chimera \cite{ic}, etc. have been reported through different models. Also, besides numerical studies, various experimental evidences for chimera states exist, such as optical coupled-map lattices \cite{exp_cml}, coupled chemical oscillators\cite{exp_co}, metronomes \cite{ic,exp_met}, and squid meta-materials \cite{exp_squid1,exp_squid2}, etc. 
	\par On the other hand,  coexisting co- and counter-rotating dynamical activity has a wide range of applications in diverse fields, including physical \cite{phy1,phy2}, biological \cite{bio,bio1}, and fluid dynamics \cite{flu1,flu2,flu3} contexts. For instance, counter-rotating spirals can be found in biological media, such as Physarum plasmodium protoplasm \cite{bio,bio1}. In physical systems, such as magnetohydrodynamics of plasma flow \cite{plasma}, and Bose-Einstein condensates \cite{BEC}, counter-rotating vortices are purposefully formed for practical reasons \cite{diba}.  As a consequence, examining the collective behaviors induced by counter-rotation becomes intriguing. The coexistence of co- and counter-rotating oscillators was initially identified by Tabor \cite{tabor}. This is followed by a series of works that have been done in co- and counter-rotating oscillators. The existence of anti-synchronization has been reported in coupled chaotic Lorentz systems \cite{as}. Subsequently, the mixed synchronization and its universal occurrence were investigated in both limit-cycle and chaotic oscillators \cite{awadhes1}. Later, the realizations of mixed synchronization were also revealed in both experimental and theoretical systems \cite{cr_exp1}. Further,  the breaking of rotational symmetry induced dynamical effects were investigated in counter-rotating oscillators with symmetry preserving and symmetry breaking couplings \cite{awadhes2}. Recently, the aging transitions have also been investigated in counter-rotating oscillators \cite{aging}. Therefore, competing co- and counter-rotating frequencies play an essential role in exhibiting a distinct set of collective behaviors including mixed synchronization and different oscillation quenching states. \textit{However, the effects of frequency and additional perturbation on the onset of the chimera states in counter-rotating oscillators are unclear and they have not been explored explicitly to the best of our knowledge.} Therefore, in this paper, we aim to investigate whether coexistence of co- and counter-rotating oscillators with additional perturbation implies chimera behavior (partial symmetry breaking dynamics)  among~ the~two groups of oscillators.
%
	\par Motivated by the above, in this paper we consider { a network of nonidentical and globally coupled Stuart-Landau oscillators} with additional perturbation. Importantly, the influence of frequencies is demonstrated by splitting the network into two groups and distributing the frequencies around a specific threshold range. To start with, the dynamical behaviors of the nonidentical counter-rotating frequencies are investigated in the absence of additional perturbation. We show that the oscillatory dynamics transits from the desynchronized state ($DS$) to the cluster oscillatory state ($COS$) via a partial synchronization ($PS$) state with increasing coupling strength. The system dynamics then attains oscillation quenching states such as amplitude death ($AD$) and oscillation death ($OD$) at larger coupling strengths. Further, we find the observed dynamical states, namely $DS$, $PS$, $COS$, and $OD$ states, break the parity$(P)$-symmetry in the absence of perturbation. When the additional perturbation is increased we find that symmetry breaking $DS$ and $COS$ become symmetry preserving incoherent mixed synchronization ($IMS$) and coherent mixed synchronization ($CMS$) states. Importantly, the symmetry breaking partial synchronization state turns into a partial symmetry broken state where a partial set of oscillators satisfies the $P-$symmetry while the others do not. {Such partial symmetry broken state   is designated here as the chimera ($CH$) state which is realized for the first time in non-identical counter-rotating oscillators. } The co-existence of symmetry preserving and symmetry breaking oscillators in the chimera state is then confirmed through the basin of attraction studies. In addition, the new measure, namely the strength of  $P-$symmetry, is used to validate the partial symmetry breaking state and its transitions.  Interestingly, we find that partial symmetry breaking occurs only when the perturbation is increased at the $DS$ and $PS$ states.{  Followed by this, the global dynamical transitions are analyzed in the parametric spaces. We can note that the increasing of the threshold increases the collective dynamical states, namely the  $DS,~ PS, ~CH, ~IMS$, and $AD$ states. In particular, the regions for the $DS,~ CH,~ IMS,~  AD$ and $OD$ states are increased with the decrease in $CMS$ and $PS$ regions. Furthermore, we also discover that increasing the strength of additional perturbation suppresses the regions of $DS,~  PS, ~ COS$ and $AD$ regions completely with the onset of $IMS,~ CH,$ and $CMS$ states.}  Finally, the robustness of the chimera is examined by increasing the size of the network at the chimera state. { Eventually, the generality of the observed oscillatory states and their transition are studied in other dynamical models such as the coupled van der Pol (vdP) and R\"ossler oscillators as well as in the phase reduced model. Notably, we find that the globally coupled vdP and R\"ossler oscillators also exhibit similar kinds of partial synchronisation and chimera states.  } \\
%
	\par The structure of the article is as follows: In Sec. II, we introduce the model of globally coupled Stuart-Landau oscillators with additional perturbation.   The dynamical transitions in the absence and in the presence of perturbation are discussed in the Secs.  III and IV, respectively. Then the mechanism and the characterization of the partial symmetry breaking state are detailed in Secs. V and VI, respectively. Further, the global dynamical transitions are discussed in Sec. VII and the robustness of the chimera state is described in Sec. VIII.   Finally, we summarize our major findings in Sec. IX. 
	\begin{figure*}[ht!]
		\includegraphics[width=17.00cm]{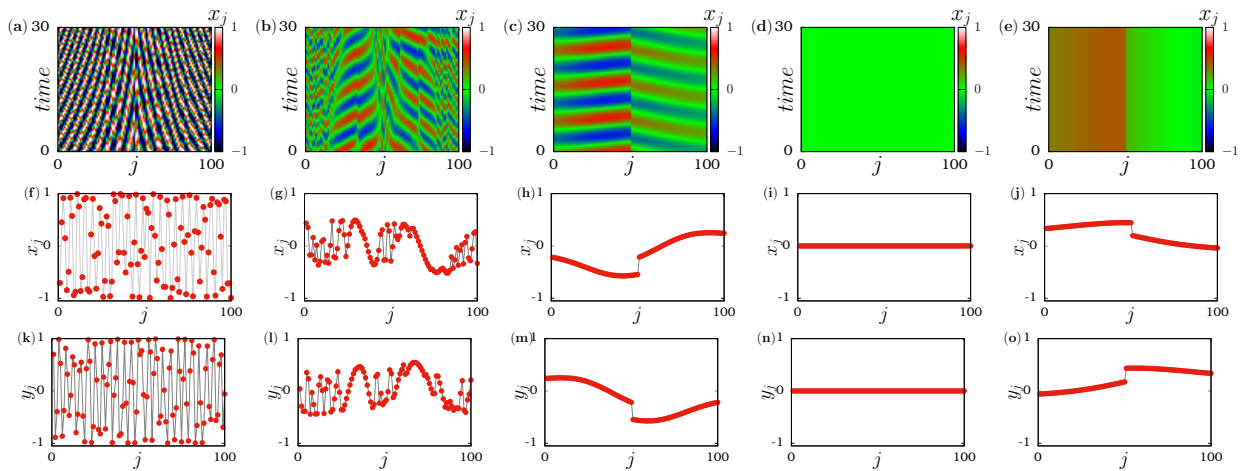}
		\caption{Spatio-temporal plots of globally coupled Stuart-Landau oscillators   distributed with nonidentical counter-rotating frequencies in the absence of perturbation ($\varepsilon=0.0$) for (a) desynchronized state ($K=0.05$), (b) partial synchronization state ($K=0.82$), (c) cluster oscillatory state ($K=1.3$), (d) amplitude death state ($K=1.86$), and (e) oscillation death state ($K=2.0$). The corresponding snapshots in terms of $x_j$ variables are displayed in (f)-(j), and the $y_j$ variables are shown in (k)-(o). Other parameters are as follows:  $\lambda=1.0$, $\Delta=0.5$, $\omega=1.0$, and $N=100$.}
		\label{dte0} 
	\end{figure*}	
	\section{ The model}
	Limit cycle oscillations  are often found in various physical, chemical, and biological phenomena including neuronal excitations, circadian rhythms, respiratory cycles, chemical oscillations, vibrations in bridges, etc. Thus, we consider a simple, prototype, self-excitatory model of Stuart-Landau (SL) limit-cycle oscillators.  Such limit cycle oscillators can be used to model a variety of nonlinear oscillators near Hopf bifurcation \cite{stu1,stu2,stu3}.  In particular, to investigate the effect of additional perturbation on non-identical counter-rotating oscillators, we consider  a system  of globally coupled Stuart-Landau oscillators whose governing equation is expressed as
	\begin{eqnarray}
	\dot{z}_j = (\lambda+ i \omega_j -|z_j|^2)z_j+ \nonumber  
	\frac{K+\varepsilon}{N}\sum_{k=1}^{N} Re({z_k}-\, {z_j})\nonumber
	\\ +i\frac{K-\varepsilon}{N}\sum_{k=1}^{N} Im ({z_k}-\, {z_j}), \quad j = 1, 2, ..., N,
	\label{model} 
	\end{eqnarray}
	where $z_j = x_j+iy_j \in \mathbb C$, $x_j$ and $y_j$ are the state variables of the SL oscillators. $\lambda$ is the Hopf bifurcation parameter and $\omega$ is the system frequency.   $K$ is the coupling strength and $\varepsilon$ is the additional perturbation in the coupling.  In the absence of additional perturbation ($\varepsilon=0$), the system preserves the rotational symmetry and  Eq. (\ref{model}) remains invariant under the gauge transformation $z_j\rightarrow z_j e^{i \theta}$.  For $\varepsilon\ne 0$, the coupled system loses this rotational symmetry. { In an earlier report, the effect of bimodal distribution was investigated in a system of globally coupled phase oscillators without time delay \cite{rev2}.   Then the influence of time-delay induced dynamics was further investigated in a large population of globally coupled phase oscillators using a bimodal frequency distribution \cite{rev1}. Further, asymmetry in bimodal frequency induced symmetry breaking has also been reported in \cite{bimodal}.  With reference to the above studies,  we partitioned the network(1) into two groups of $N/2$ oscillators each (sub-populations) designated as $z_j^{(1)}$ and $z_j^{(2)}$ and the corresponding frequencies are $\omega_j^{(1)}$ and $\omega_j^{(2)}$, respectively.   Specifically, the first group is distributed with a  uniform frequency distribution in the range $\omega_j^{(1)}=(\omega - \Delta, ~ \omega + \Delta)$ for $j=1, 2, ..., N/2$, while the second  group is distributed uniformly in the frequency range $\omega_j^{(2)}=(-\omega - \Delta, ~ -\omega + \Delta)$ for $j=N/2+1, ..., N$.}  Depending on the positive and negative signs infront of $\omega$, the individual oscillator rotates either in the  clockwise or anticlockwise direction.  
	For numerical simulations, we used the Runge-Kutta $4^{th}$ order scheme with a step size of $h=0.01$. 
	\section{Dynamical transitions in the absence of additional perturbation ($\varepsilon=0.0$) and parity symmetry}
		\begin{figure*}[htb!]
		\includegraphics[width=18.00cm]{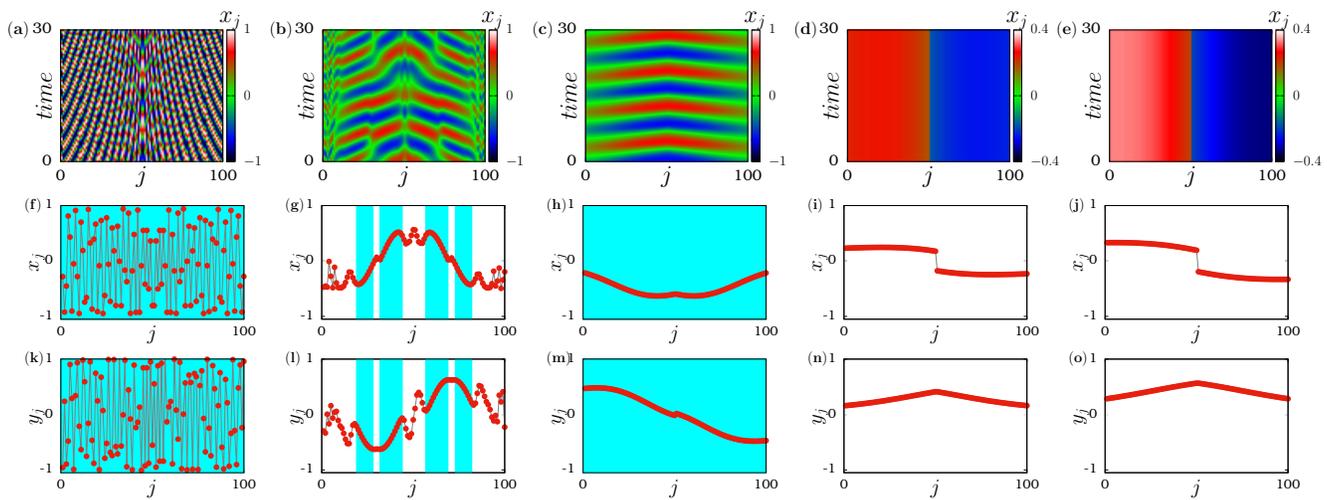}
		\caption{Spatio-temporal plots of globally coupled Stuart-Landau oscillators distributed with nonidentical counter-rotating frequencies in the presence of perturbation ($\varepsilon=0.1$) for (a) incoherent mixed synchronization state ($K=0.05$),  (b) chimera  state ($K=0.82$),  (c)  coherent mixed synchronization state ($K=1.3$),  (d) oscillation death state ($K=1.86$), and (e) oscillation death state ($K=2.0$). The shaded areas in Figs.\ref{dte0.1}(f)-\ref{dte0.1}(h) and  Figs.\ref{dte0.1}(k)-\ref{dte0.1}(m) illustrate the $P-$symmetry preserving oscillators.  The corresponding snapshots in terms of $x_j$ variables are displayed in  figures (f)-(j) and the $y_j$ variables are shown in figures (k)-(o). Other parameters are as follows: $\lambda=1.0$, $\Delta=0.5$, $\omega=1.0$, and $N=100$.}
		\label{dte0.1} 
	\end{figure*}
	At first, we examine the dynamical transitions in globally coupled SL oscillators in the absence of additional perturbation ($\varepsilon=0.0$). To show the dynamical transitions, the spatio-temporal behavior (in the upper panel) and the corresponding snapshots of $x_j$ and $y_j$ variables (in the middle and the lower panels, respectively)  are plotted  in Fig.~\ref{dte0} by fixing $\Delta=0.5$, $\omega=1.0$  and for different values of coupling strength ($K$).  If the coupling strength is low (for $K=0.05$), we notice the existence of a desynchronized state ($DS$)  as in Fig.~\ref{dte0}(a).  In the desynchronized state,   all the oscillators in the first and second groups are distributed randomly as evident from the snapshots in  Figs.~\ref{dte0}(f) and \ref{dte0}(k). { When the coupling strength is increased to $K=0.82$, there exists the coexistence of coherent and incoherent behaviors (see Fig.~\ref{dte0}(b)) in each of the groups and it is further referred to as partial synchronization  ($PS$) state \cite{ps1,ps2,ps3}.}  Co-existence of  coherent and incoherent patches in the partial synchronization state is also clear from the snapshots of  $x_j$ and $y_j$ variables in Figs.~\ref{dte0}(g) and \ref{dte0}(l).   Upon increasing the coupling strength to  $K=1.3$, we find the emergence of cluster oscillatory state ($COS$).  Here we observed two cluster states, where  the first group (co-rotating oscillators) forms one group and the corresponding  second  set (counter-rotating oscillators) forms the other group as shown in Figs.~\ref{dte0}(c), \ref{dte0}(h),  and \ref{dte0}(m). { In particular,  individual clusters exist as a travelling wave, and on comparing both the clusters (of $y_j$ variables), we find that  they  are antiphase with each other \cite{rev1,rev5,io}. } Increasing the coupling strength further ($K=1.86$), all the oscillators in the network attain the homogeneous steady state (see Fig.~\ref{dte0}(d))  namely amplitude death ($AD$) state.   As a result,  all the oscillators in the network acquire the trivial steady state in the $x_j$ and $y_j$ variables (see Figs.~\ref{dte0}(i),  and \ref{dte0}(n)).    Further, at even larger coupling strength ($K=2.0$), the oscillators in the network split into two groups as in the cluster oscillatory state and populate into two different inhomogeneous steady states constituting the oscillation death ($OD$) state which is shown in Figs.~\ref{dte0}(e), \ref{dte0}(j) and \ref{dte0}(o).   Thus, it is clear that the transitions from desynchronized state to cluster oscillatory state through the partial synchronization state occur first as the coupling strength increases. Then, the transition to oscillation quenching states such as amplitude death and oscillation death states are observed at larger coupling strengths. 
	\par  	On the other hand, it has been revealed from the earlier reports that the pair of counter-rotating oscillators preserve a set of permutation and permutation-parity symmetries in the $x_j$ and $y_j$ variables of the same system. Such combined symmetry  preservation causes  mixed synchronization among the oscillators \cite{awadhes1}. In our case, according to the distribution of counter-rotating  frequencies, the considered network is designed to have the following symmetries in the $x_j$ and $y_j$ variables, 
	\begin{eqnarray}
	x_j{^{(1)}}=x_{N-j+1}{^{(2)}},  \nonumber \\
	y_j{^{(1)}}=-y_{N-j+1}{^{(2)}}.
	\end{eqnarray}
	Here the superscripts ${(1)}$ and ${(2)}$ stand for the groups corresponding to co- and counter-rotating oscillators, respectively. We simply call this symmetry as parity symmetry or $P-$symmetry throughout the text.  
	However, by comparing the snapshots of $x_j$ and $y_j$ variables in Fig.\ref{dte0}, we notice that the observed dynamical states do not preserve the above mentioned $P-$symmetry in the absence of additional perturbation.
{ Furthermore, heterogeneity or asymmetry in time-delay, frequency, and coupling strength in globally coupled oscillators has been extensively  investigated \cite{rev3,rev4,rev5}.  For instance,  heterogeneity in time-delay induced synchronization has been described in \cite{rev3}. Then the  asymmetry in frequency and coupling strength associated with macroscopic traveling waves has also been studied in the Kuramoto model \cite{rev4}. Motivated by these studies, we implement asymmetry in the coupling via additional perturbation to analyze the symmetry of the dynamical states.} Thus, in the following,  we investigate the dynamical transitions for the same set of parameters with an additional perturbation ($\varepsilon \ne 0$ in Eq.~(\ref{model})), to  understand  whether the observed symmetry breaking dynamical states can preserve  the  $P-$symmetry  due to the additional perturbation.
	\section{Dynamical transitions in the presence of additional perturbation ($\varepsilon=0.1$)}
	\par  In order to understand the effect of the additional perturbation on the dynamical states, we plotted the spatio-temporal behavior and the corresponding snapshots of $x_j$ and $y_j$ variables  in  Figs.~\ref{dte0.1}  by fixing the strength of the additional perturbation as $\varepsilon=0.1$. The shaded areas in the snapshots in Figs.\ref{dte0.1}(f)-\ref{dte0.1}(h) and Figs.\ref{dte0.1}(k)-\ref{dte0.1}(m)  denote the regions containing the $P-$symmetry preserving oscillators.  
	For $K=0.05$, we observed the incoherent mixed synchronization $(IMS)$ state as in  Fig.~\ref{dte0.1}(a). Such an incoherent mixed synchronization state preserves the  $P-$symmetry as described below. The first and second groups preserve the symmetry $x_j{^{(1)}}=x_j$, and  $x_{N-j+1}{^{(2)}}=x_j$  in  $x_j$ variables  as clear from Fig.~\ref{dte0.1}(f).  Simultaneously,  the corresponding   $y_j$ variables of the first and second groups  preserve $y_j{^{(1)}}=y_j$, and  $y_{N-j+1}{^{(2)}}=-y_j$  symmetry  as shown in Fig.~\ref{dte0.1}(k). Though the oscillators are having random values in the  $x_j$ and $y_j$  variables, ~the oscillators in the network preserve the above  $P$-symmetry.  Hence the state in Figs.\ref{dte0.1}(a) or \ref{dte0.1}(f)  and \ref{dte0.1}(k)  is referred to as an incoherent mixed synchronization $(IMS)$ state. By increasing the coupling parameter to $K=0.82$,  we found coexisting coherent and incoherent patches in each group as seen in Fig.~\ref{dte0.1}(b). { Compared to the partial synchronization state which completely breaks the $P-$symmetry (see  Figs.~\ref{dte0}(b),(g),(l)),  here we notice that the oscillators in the coherent patches of the first and second groups preserve the $P-$symmetry in $x_j$ and $y_j$ variables. As a consequence, coexisting  symmetry preserving and symmetry breaking oscillators are noticed in Figs.~\ref{dte0.1}(g) and Fig.~\ref{dte0.1}(l). The shaded regions correspond to symmetry preserving oscillators while the unshaded regions represent the symmetry breaking oscillators. Such a ~partial symmetry breaking state is interesting and it is referred to as chimera $(CH)$ state here. Thus, even though the oscillators in the network exhibit spatially coexisting coherent and incoherent behaviors, if they completely break the  $P-$symmetry or completely preserve the $P-$symmetry, we referred to such a dynamical state as partial synchronization; otherwise, if the coherent and incoherent patches are partially symmetry broken, the state is designated as a chimera.  We believe that this kind of chimera state in nonidentical counter-rotating oscillators has been observed for the first time and it has not yet been realized in the earlier literature on counter-rotating oscillators.}  When  $K=1.3$, all the oscillators in both the first and second groups preserve the $P-$symmetry.  As a result, we find a coherent cluster oscillatory state  which is further represented as a coherent mixed synchronization state $(CMS)$.  The coherent mixed synchronization state among the two groups is clearly depicted in  Figs.~\ref{dte0.1}(c),~\ref{dte0.1}(h), and \ref{dte0.1}(m).  Further, the additional perturbation suppresses the amplitude death state.  As a result, we observe oscillation death alone at the larger coupling strengths $\varepsilon=1.86$ and $\varepsilon=2.0$. From Figs.~\ref{dte0.1}(d), \ref{dte0.1}(i),  \ref{dte0.1}(n) and Figs.~\ref{dte0.1}(e), \ref{dte0.1}(j),  \ref{dte0.1}(o), it is clear that the observed oscillation death states are symmetry breaking.  Therefore, we observed the transition from incoherent mixed synchronization $(IMS)$  to coherent mixed synchronization $(CMS)$ through chimera $(CH)$  state and finally to oscillation death $(OD)$ state when the additional perturbation is included. 
	\par { The above observed oscillatory patterns are confirmed further using the phase reduced model described in Appendix A. }
	\section{Mechanism for partial symmetry breaking}	
	\par
\begin{figure}[ht!]
		\includegraphics[width=8.00cm]{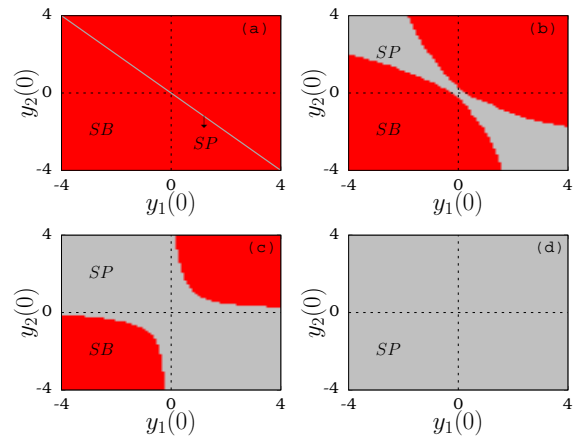}
		\caption{ Basin of attraction for two coupled oscillators by fixing the  state variables $x_1(0)=0.9$ and $x_2(0)=0.9001$ and by varying the state variables $y_1(0)$ and $y_2(0)$ for  different perturbation strengths, (a) $\varepsilon=0.0$,  (b) $\varepsilon=0.0025$, (c) $\varepsilon=0.003$,  and (d) $\varepsilon=0.1$.  The grey and red colors represent the basin for symmetry preserving state and symmetry breaking state, respectively. Other parameters: $\lambda=1.0$, $K=1.0$,  $\omega=1.0$ and $N=2$. }
		\label{basin} 
	\end{figure}

  In order to understand the mechanism for the co-existence of multistable states, i.e. coexisting partial symmetry preserving and symmetry breaking states, we have performed a basin stability analysis in the following.  Due to multistability, a partial set of oscillators in the network alone preserves the $P-$symmetry, resulting in the coexistence of symmetry preserving and symmetry  breaking behavior in the dynamical states. 	In order to show the role of  multistability among the symmetry preserving and symmetry breaking dynamical states in the initial state space, we consider a minimal network of two coupled Stuart-Landau oscillators with counter-rotating frequencies.  The corresponding    basin of attraction is depicted  in  Fig.~\ref{basin}, by fixing $x_1(0) =0.90$ and $x_2(0)=0.901$ and by varying the state variables  $y_1(0)$ and $y_2(0)$ for $K=1.0$.  Figure \ref{basin}(a) is plotted for the absence of perturbation, namely $\varepsilon=0.0$.  We can note that the system dynamics preserves $P-$symmetry only for a specific initial state (i.e. when the initial states are equal and opposite),  otherwise the initial state space is filled with a symmetry-breaking state.
	\begin{figure*}[htb!]
	\includegraphics[width=18.00cm]{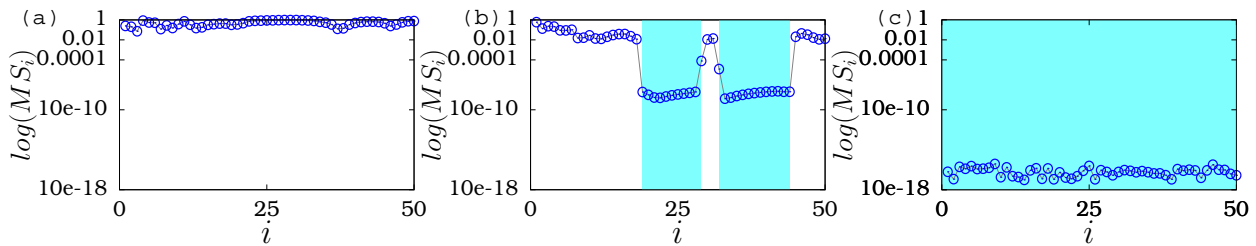}
	\caption{Mixed synchronization error estimated using Eq. (\ref{MSerr}) at coupling strength $K=0.82$ for specific perturbation strengths (a) $\varepsilon =0.0$,  (b) $\varepsilon=0.1$, and (c) $\varepsilon=0.2$.     }  
	\label{er1} 
\end{figure*} 	
		The symmetry preserving state observed in Fig.~\ref{basin}(a) is unstable and it exists only for  specific asymmetric initial conditions. Here, the asymmetric initial condition corresponds to  the case where  $y_1(0)$  is positive while $y_2(0)$  is negative or vice versa.     When  the additional perturbation in the coupling is increased to $\varepsilon=0.0025$, we find from the asymmetric initial state that either $y_1(0)$ or $y_2(0)$ positive while the other one is negative favors the symmetry preserving behavior and it is clearly depicted in Fig.~\ref{basin}(b).  As a result, we find that for some initial conditions, the system dynamics preserves  $P-$symmetry, while for others, it breaks $P-$symmetry. In this manner, the coexistence of symmetry preserving and symmetry breaking states are realized. Upon increasing the perturbation further to $\varepsilon=0.003$, an increase in the symmetry preserving state basin is observed as in Fig.~\ref{basin}(c). On increasing the perturbation further one finds that the entire basin is occupied by a symmetry preserving state (see Fig.~\ref{basin}(d)).  Thus, from the basin stability analysis, one can confirm the presence of multistability among the  symmetry preserving  and   symmetry breaking  states. Because of the multistable nature, we find the partial symmetry breaking states (co-existence of symmetry preserving and symmetry breaking dynamical behavior) in the larger network.  In the following section,  we discuss the characterization of the partial symmetry breaking state using the notion of the strength of $P-$symmetry in some detail.   
	\section{Characterization of partial symmetry breaking states}
	\par  
	In order to check whether the observed dynamical states preserve $P-$symmetry, we first measure the error of $P-$symmetry preservation by finding the difference and sum of  $x_j$  and $y_j$ variables which  can be expressed as 
	\begin{eqnarray}
	\label{MSerr}
	PS_{i} = \sqrt{(x_j-x_{N-j+1})^2+(y_j+y_{N-j+1})^2},\\
	j=1,2,..., N,~   ~ i=1,2,..., N/2, \nonumber
	\end{eqnarray}
	where $PS_i$ is the parity symmetry error for the $i^{th}$ pair. 
	Note that here the index $i$ represents the $P-$ symmetry error between the pair of first and second sub-populations. If the oscillators in the first and second sub-populations preserve the $P-$symmetry, the error takes the null value,  otherwise, the system does not preserve the $P-$symmetry.  The error estimated from Eq.~(\ref{MSerr}) is plotted in Fig.~\ref{er1}  for $K=0.82$.  In the absence of perturbation $\varepsilon=0.0$, we find that the system does not preserve  $P-$symmetry (see Fig.~\ref{er1}(a)),  since all the $PS_i$'s take the values between $0<PS_i\le 1$.  On~ increasing the perturbation, one finds that a partial set of oscillators take  null values while the remaining set takes non-zero values (see Fig.~\ref{er1}(b)). Symmetry preserving oscillators are represented by the shaded regions.   As a consequence, it is observed that the partial symmetry breaking state   (coexistence of partial symmetry preserving and partial symmetry breaking dynamical states) exhibits the error in synchronization as in Fig.~\ref{er1}(b).    Upon increasing the perturbation strength further it is observed that all the oscillators are showing zero error, that is  the system dynamics completely preserves  $P-$symmetry as depicted in Fig.~\ref{er1}(c).  Thus, we observe the transition from symmetry breaking state to symmetry preserving state through partial symmetry breaking state.
	\par  In addition, to quantify $P-$symmetry in each dynamical state and their transition as a function of  parameters, we introduced a new measure that quantifies the strength of the $P-$symmetry. The  corresponding expression for the strength of $P-$symmetry  is  
	\begin{eqnarray}
	S_{PS}= \frac{\sum_{i=1}^{N/2} H_{PS_i}}{N/2},  \qquad H_{PS_i}=\Theta(\delta-PS_i)
	\label{sms1}
	\end{eqnarray}
	where $\delta$ is a small threshold (which is taken as $0.001$ for this study),  $\Theta$ is the Heaviside step function, and  $PS_i$ is the error of $P-$symmetry.  	If $PS_i$ is less than $\delta$,   $H_{PS_i}$  takes the value unity, otherwise, it is zero.  Then the measure estimates the ratio of $P-$symmetry  in each dynamical state. If the strength of the $P-$symmetry $S_{PS}$ is unity, the dynamical state completely preserves the $P-$symmetry and if it is zero then it represents a completely symmetry broken state.  The  strength of the $P-$symmetry between~ $0<S_{PS}<1$ represents the partial symmetry broken state.   
	\par
\begin{figure}[htb!]
	\includegraphics[width=9.00cm]{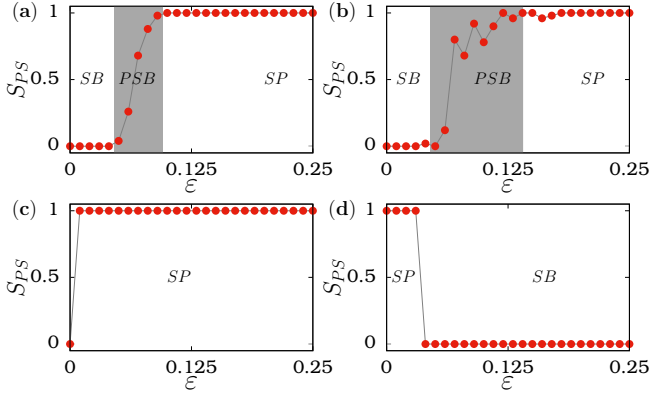}
	\caption{ The strength of mixed synchronization symmetry $(S_{PS})$  as a function of perturbation strength $\varepsilon$ by fixing the coupling parameters at (a) desynchronized state ($K=0.05$),  (b) partial synchronization state ($K=0.82$),  (c) cluster oscillatory state ($K=1.3$), and  (d)  amplitude death state ($K=1.86$).    $SB$,  $PSB$ and $SP$ are the symmetry breaking, partial symmetry breaking and symmetry preserving dynamical transitions, respectively.  Other parameters: $\lambda=1.0$, $\Delta=0.5$, $\omega=1.0$, and $N=100$. }  
	\label{sms} 
\end{figure}

 To understand the effect of additional perturbation on  different dynamical states as observed in Figs.~\ref{dte0} and \ref{dte0.1}, we have plotted the strength of $P-$symmetry  ($S_{PS}$) as a function of perturbation ($\varepsilon$)  in Fig.~\ref{sms}. The filled circles represent the range of  the strength of $P-$symmetry ($S_{PS}$) corresponding to the additional perturbation $\varepsilon$.  Figure \ref{sms}(a) is plotted for the dynamical transition at the desynchronized state (for  $K=0.05$) which shows that $S_{PS}$ is zero in the absence of perturbation indicating a symmetry broken ($SB$) state. By increasing the perturbation, the symmetry broken state transits to a symmetry preserving ($SP$) state through partial symmetry broken ($PSB$) state.  In the partial symmetry broken  state the range of $S_{PS}$ takes the  value between $0<S_{PS}<1$ and the unit value of $S_{PS}$ indicates the maximum strength of $P-$symmetry which is a symmetry preserving state.  The shaded region denotes the partial symmetry broken state. A similar transition is also observed in the partial synchronization state plotted in Fig.~\ref{sms}(b) for $K=0.82$. In comparison with Fig.~\ref{sms}(a), the region for $PSB$ state is seen to widen in Fig.~\ref{sms}(b).  Further,  in the cluster oscillatory state, initially, the system  is in a symmetry breaking state then it directly transits to symmetry preserving state by slightly increasing the perturbation strength (see Figure \ref{sms}(c) for $K=1.3$).  In contrast,  by fixing the parameter in the amplitude death state and increasing the perturbation, the value of  $S_{PS}$ transits from $1$ to $0$ as a function of perturbation, that is the transition from symmetry preserving to symmetry breaking (see Fig.~\ref{sms}(d)) state occurs.  Importantly, here, the observed symmetry breaking  state is the oscillation death state.  Furthermore, increasing the perturbation at the oscillation death state always shows a null value of  $S_{PS}$ (which has not been displayed here) which indicates the symmetry breaking state for all values of further perturbation. Thus, from the observations, it is to be noted that partial symmetry breaking exists only in the originally desynchronized and partial synchronization states. Next, the global dynamical transitions are delineated as a function of frequency thresholds  and  perturbation strengths in the following.  
	\section{Global dynamical transitions}	
\begin{figure}[ht!]
		\hspace{-0.5cm}
		\includegraphics[width=9.00cm]{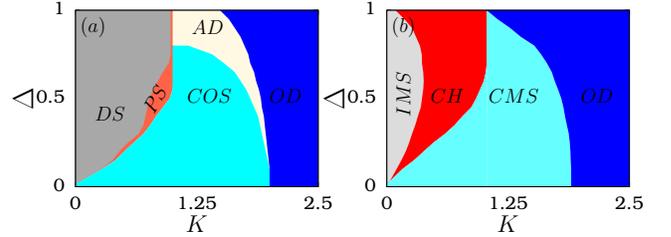}
		\caption{Two parameter diagrams of globally coupled nonidentical counter-rotating  oscillators in $(K, \Delta)$ space with different perturbations,  (a) $\varepsilon = 0.0$, and (b)  $\varepsilon = 0.1$.  $DS$,  $PS$, and $COS$ are the desynchronized state, partial synchronization state and cluster oscillatory state, respectively. $IMS$, $CH$ and $CMS$ are the incoherent mixed synchronization, chimera  and coherent mixed synchronization states, respectively.    $AD$ and $OD$ correspond to the amplitude death and oscillation death states, respectively. Other parameters: $\lambda=1.0$,  $\omega=1.0$, and $N=100$. }
		\label{dtr} 
	\end{figure}

	The global dynamical transitions are analyzed by using the measure of  the strength of incoherence as discussed in \cite{sic1,sic2} as  well as the strength of $P-$symmetry. Primarily, to show the  dynamical transitions  in the parametric space, we have plotted the two-parameter diagrams in  the $(K, \Delta)$ space for  $\omega=1.0$.  Figure~\ref{dtr}(a) is plotted for the absence of perturbation ($\varepsilon = 0.0$), and it clearly depicts that at lower values of threshold, there exists a direct transition from desynchronized state to cluster oscillatory state.   The region for the desynchronized state becomes increased with an increase in the threshold. Further, a moderate value of threshold exhibits  the transition from desynchronized state to cluster oscillatory state through the partial synchronization state.  On increasing the threshold further, the region for  cluster oscillatory state gets decreased and suppressed completely at  larger thresholds. At larger coupling strengths, we find that the existence of oscillation death state is confirmed and that the region for oscillation death state gets increased with the increase in the threshold.  In addition, while increasing the threshold, we notice a new additional branch of amplitude death state between the cluster oscillatory state and oscillation death state.  	
	Furthermore,  Fig.~\ref{dtr}(b)  is plotted for the additional perturbation $\varepsilon=0.1$.  At lower values of coupling strength, we observe the incoherent mixed synchronization state, and the region for incoherent mixed synchronization state is increased up to a certain threshold and then it again decreases.  As a function of coupling strength,  the incoherent mixed synchronization state transits to  coherent mixed synchronization state through the chimera state (partial symmetry breaking state). The region for chimera state increases with a decrease in coherent mixed synchronization state region. Further,  oscillation death state is noticed at still larger coupling values and the spread of the $OD$ region gets increased as a function of threshold.  Also, we find that the additional perturbation suppresses the amplitude death state completely.	
	. 
	\begin{figure}[htb!]
		\hspace{-0.5cm}
		\includegraphics[width=9.00cm]{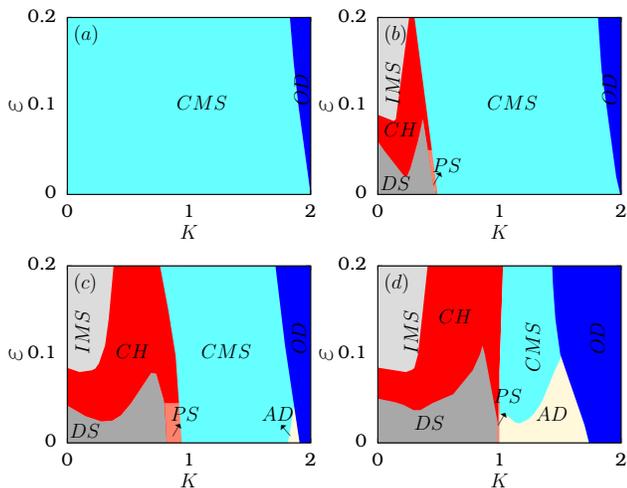}
		\caption{Two parameter diagrams of globally coupled nonidentical counter-rotating oscillators in $(K,\varepsilon)$ parametric space by fixing different thresholds for (a) $\Delta=0.0$, (b) $\Delta=0.2$, (c) $\Delta=0.5$, and (d) $\Delta=0.8$.   Other parameters are same as in Fig.~\ref{dtr}}
		\label{pert} 
	\end{figure} 
	\par   Additionally, the two-parameter diagrams are displayed in the $(K, \varepsilon)$ parametric space in Fig.\ref{pert} for four distinct values of the threshold  to provide a better understanding of the dynamical transitions as a function of the perturbation.   Figure \ref{pert}(a) is depicted for the identical counter-rotating frequencies,  i.e. by fixing the threshold $\Delta=0.0$. We discover that the identical counter-rotating oscillators display a direct transition from $CMS$ to $OD$ and that the region for $CMS$ state decreases with increasing $OD$ region.  By increasing the threshold to $\Delta=0.2$, we find that the onset of distinct dynamical states at the lower coupling strengths (see Fig.~\ref{pert}(b)).  Importantly, we find the $DS$ state at the lower values of coupling strength and perturbation.  Then, it transits to $IMS$ state through $CH$ state while increasing the perturbation. In the absence of perturbation, increasing the coupling exhibits a transition from $DS$  state to $OD$ state via  $PS$ and $COS$ states (not explicitly displayed in the two parameter diagrams). Further, when the perturbation strength is slightly increased, the $COS$ state becomes a $CMS$ state. Comparing Fig.~\ref{pert}(a), the $CMS$ state region is shortened  in Fig.~\ref{pert}(b). At larger coupling, there exists the $OD$ region and it increases as a function of perturbation.  On increasing the value of the threshold  even more to $\Delta=0.5$ and $\Delta=0.8$,  we notice that the regions for $DS$,  $CH$, $IMS$, and $OD$ states get increased with decreasing  $CMS$ region.  Additionally, we notice the onset of a new branch of amplitude death state at lower values of perturbation before the onset of oscillation death state, and its region is enlarged at a larger threshold. In addition, the AD region  vanishes completely while increasing the perturbation to  larger values.  In the next section, we examine whether the observed chimera is robust in larger network sizes.    
	\section{Robustness of chimera at  larger sizes of the network}
	\begin{figure}[h!]
		\vspace{0.1cm}
		\includegraphics[width=9.00cm]{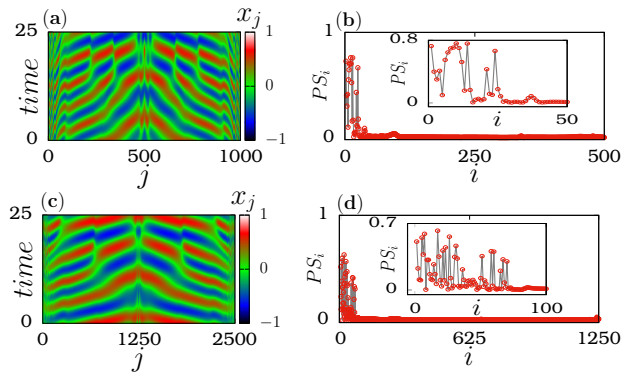}
		\caption{Spatio-temporal behavior and error of $P-$symmetry for two groups of non-identical counter-rotating  oscillators for the network size (a)-(b) $N=1000$ and (c)-(d) $N=2500$ by fixing the coupling strength $K=0.82$.  The inset in (b) and (d) denote zoomed views of the symmetry breaking oscillators in the chimera state. Other parameters: $\lambda=1.0$, $\Delta=0.5$, $\omega=1.0$, and $\varepsilon=0.1$.  }
		\label{large} 
	\end{figure}
	\par The robustness of the observed chimera state is also investigated at $K=0.82$, using spatio-temporal behavior and the corresponding   error of $P-$symmetry in Fig.~\ref{large} by fixing the network sizes as $N=1000$ and $N=2500$. From the spatio-temporal plots in Figs.~\ref{large}(a) and \ref{large}(c), it is clear that the existence of the chimera state with coexisting coherent and incoherent patches of oscillators for the network sizes $N=1000$ and $N=2500$. Further, the partial symmetry breaking in the chimera state is also confirmed through the error of $P-$symmetry measure in Figs.~\ref{large}(b) and \ref{large}(d) which clearly depict that some of the oscillators are symmetry breaking ($PS_i >0$) while the remaining oscillators are symmetry preserving ($PS_i =0$). The inset of Figs.~\ref{large}(b) and \ref{large}(d) denote the zoomed view of the symmetry-breaking oscillators.  Thus, it is clear that the structure of the observed chimera state is also unaffected by increasing the size of the network and it is independent of the number of the oscillators in the network.
	\par { In addition, the generality of the above observed dynamical states and their transitions will be confirmed for two other different dynamical models, namely globally coupled van der Pol and R\"ossler oscillators in Appendix B.  We will explicitly demonstrate the existence of partially synchrnonized as well as chimera states in these systems also to prove their generic nature.}
	\section{Conclusions}
	\par Co-existence of co- and counter-rotation can be found in many physical, mechanical, and biological systems. In particular, frequency plays a significant role in generating co- and counter-rotation in a dynamical system. Hence, in the present work, we have explored the dynamical transitions by distributing the non-identical counter-rotating frequencies.{  To exemplify this, we considered a network of non-identical and  globally coupled Stuart-Landau oscillators with additional perturbation.} Primarily, the dynamical transitions are investigated in the absence of perturbation and it is observed that  transitions from desynchronized state to cluster oscillatory state via the partial synchronization state in the oscillatory regime and amplitude death to oscillation death in the oscillation quenching regime do occur. Further, we find that the observed dynamical states are breaking the parity$(P)-$symmetry in the absence of perturbation. When the perturbation is increased the system preserves $P-$symmetry and exhibits incoherent and coherent mixed synchronization states. Interestingly, we have identified the partial symmetry breaking state between the incoherent and coherent mixed synchronization states.{ Such a partial symmetry breaking state is referred to as a chimera state. } { Thus, when the entire set of oscillators in the coherent and incoherent groups either preserve or break the $P-$symmetry as a whole, then it is referred to as the partial synchronization state. Otherwise, if only a partial set of the oscillators break  the $P-$symmetry while the remaining oscillators preserve the $P-$symmetry then it is called as the chimera state. } The mechanism for the coexistence of partial symmetry breaking and symmetry preserving dynamics in the chimera state is further confirmed through the basin stability analysis. Followed by this, we introduced a statistical measure, namely   the strength of $P-$symmetry, to quantify the partial symmetry breaking state. Using this measure, the existence of the partial symmetry breaking state is validated and it exists only when fixing the coupling parameter at the desynchronized and partial synchronization states. Moreover, the global dynamical transitions are also investigated in the presence and in the absence of perturbation as well as by fixing different frequency thresholds. Increasing the threshold enlarges the regions for desynchronized state, incoherent mixed synchronization, chimera, and oscillation death states. Ultimately, raising the threshold range decreases the region of coherent mixed synchronization state. The onset of amplitude death is also observed for increasing the thresholds and it gets suppressed while increasing the perturbation. Finally, we have also examined the robustness of the chimera state by increasing the network size. We find that the observed chimera state is structurally similar and robust at larger sizes of the network as well.  { In summary, we have revealed for the first time the occurrence of novel chimera states in a network of globally coupled non-identical counter-rotating Stuart-Landau oscillators in this work. We further examined the generality of the observed oscillatory states and their transitions in other dynamical systems as well.  It is discovered that the newly observed partial synchronization and chimera states, as well as their dynamical transitions, are robust in globally coupled van der Pol and R\"ossler oscillators also.} We believe that our findings will shed further light on the dynamical transitions in counter-rotating oscillators in the science and engineering disciplines.	  
	\section*{Acknowledgments} 
	\par KS thank the DST-SERB, Government of India, for providing National Post Doctoral fellowship under the Grant No. PDF/2019/001589. The work of VKC is supported by the SERBDST-MATRICS Grant No. MTR/2018/000676 and DSTCRG Project under Grant No. CRG/2020/004353 and VKC wish to thank DST, New Delhi for computational facilities under the DST-FIST programme (SR/FST/PS- 1/2020/135) to the Department of Physics. M.L. wishes to thank the Department of Science and Technology for the award of DST-SERB National Science Chair. 	
	\appendix
	\section{ Phase reduced model: Co-existence of coherent and incoherent dynamical behaviors and their dynamical transitions}
	In addition to the results discussed in the main text, the observed oscillatory states are validated here in this Appendix using a reduced phase model, which can be derived by substituting    $z_j=r_j e^{i\theta_j}$ in Eq.~(\ref{model}). Then by separating the real and the imaginary parts and by considering $r_j=r_k=1$, we can get the reduced phase model expression  as given below
	\begin{eqnarray}
	\dot{\theta_j}= \omega_j -\frac{K+\varepsilon}{N}\sum_{k=1}^{N} (\cos(\theta_k)-\cos(\theta_j))\sin{\theta_j} \nonumber  \\ +\frac{K-\varepsilon}{N}\sum_{k=1}^{N} (\sin(\theta_k)-\sin(\theta_j))\cos{\theta_j} . 
	\label{aeq2} 
	\end{eqnarray}
	\begin{figure}[h!]
		\vspace{-0.0cm}
		\includegraphics[width=9.0cm]{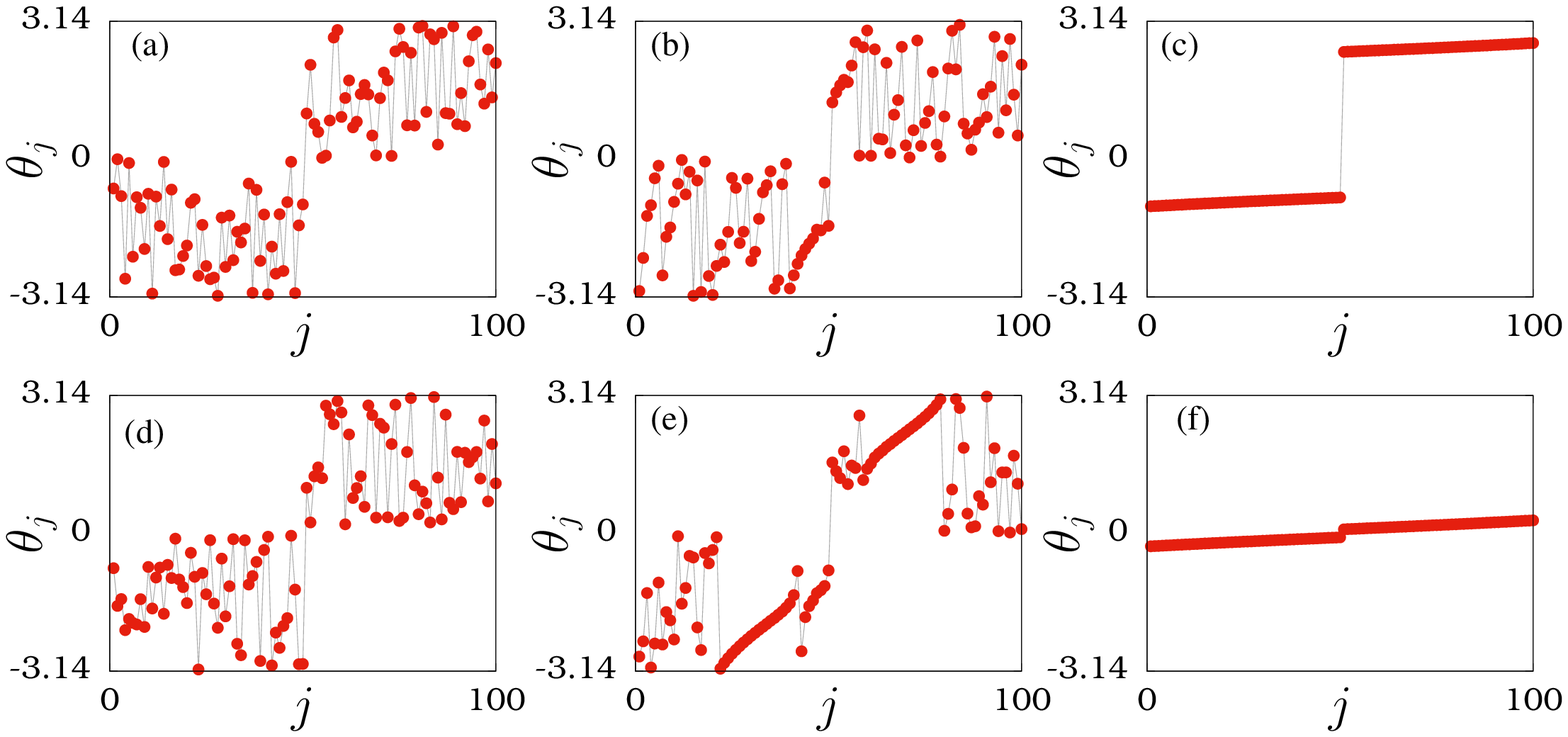}
		\vspace{-0.3cm}
		\caption {Phases of oscillators are plotted using  Eq.~(\ref{aeq2}) and by fixing $\varepsilon=0.0$ for (a) $K=0.1$ (desynchronized state), (b)  $K=0.65$ (partial synchronization state), and    (c) $K=1.5$ (cluster oscillatory state).  (d)-(e) are the corresponding dynamical behaviors at the perturbation strength $\varepsilon=0.3$. Other parameters: $\Delta=0.5$, $\omega=1.0$, and $N=100$. } 
		\label{fphase}
	\end{figure}
	\begin{figure}[h!]
	\centering
	\vspace{0.1cm}
	\includegraphics[width=9.00cm]{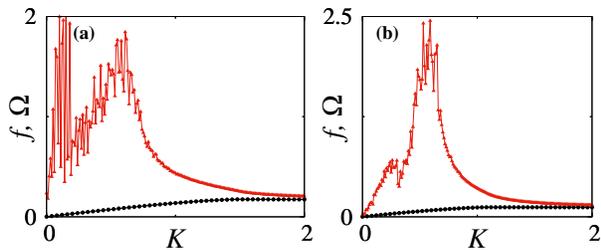}
	\caption{Mean-field ($\Omega$) and mean-ensemble frequency ($f$) as a function of coupling strength ($K$) for (a) $\varepsilon=0.0$ and (b)   $\varepsilon=0.3$. Red and black linepoints indicate the range of mean-field and mean-ensemble frequencies. Other parameters are same as in Fig.~\ref{fphase}. }
	\label{mean} 
\end{figure}	
	Using Eq.~(\ref{aeq2}), the phases of the oscillators are plotted in Fig.~\ref{fphase}. Figures~\ref{fphase}(a)-\ref{fphase}(c) in the top panel correspond to the absence of perturbation, that is for $\varepsilon=0.0$, and Figs. \ref{fphase}(d)-\ref{fphase}(f) in the lower panel depict the dynamical states for the perturbation strength $\varepsilon=0.3$. In Fig.~\ref{fphase}(a) for $K=0.1$, we can see that the phases of the oscillators are randomly distributed, with the first group (co-rotating oscillators) having negative random phases and the second group (counter-rotating oscillators) showing positive random phases. Such a state with  incoherent random phases corresponds to a desynchronized state. When the coupling strength is increased to  $K=1.1$, some of the oscillators in the first and second groups form coherent patches with coherent phases while the remaining oscillators portray incoherent phases, as shown in Fig.~\ref{fphase}(b). Such coexistence of coherent and incoherent phases in a dynamical state in globally coupled oscillators is known as partial synchronization.  On increase of the coupling strength to even higher value (for example, $K=2.0$), the phases of co- and counter-rotating oscillators display two independent clusters resulting in the cluster oscillatory state as shown in Fig.~\ref{fphase}(c). Thus, it is clear that the phases of the oscillators also show a dynamical transition from the desynchronized state to the cluster oscillatory state through the partial synchronization state.

	\par In addition, the oscillatory states are also analyzed in the presence of  the additional perturbation by fixing $\varepsilon=0.3$. From  Figs.~\ref{fphase}(d)-\ref{fphase}(f), we can note that the system~(\ref{aeq2}) exhibits a similar dynamical bahaviors even in the presence of additional perturbation as  seen in Figs. ~\ref{fphase}(a)-\ref{fphase}(c). However, in this case, all of the oscillators in Figs.~\ref{fphase}(d) show symmetry breaking, resulting in a desynchronized state. It is to be noted that the coherent patches in Fig.~\ref{fphase}(e) are symmetry preserving while the incoherent patches are symmetry breaking.  Therefore Fig.~\ref{fphase}(e) is a partial symmetry broken state that is analogous to a chimera state. Similarly, the coherent patches in Figs.~\ref{fphase}(f) preserve symmetry, hence it is equivalent to incoherent mixed synchronization.  In this manner, we validated our results using the phase-reduced model as well.  Particularly, the existence of partial synchronization and chimera states have been confirmed.       
	\par 
	
	  Furthermore, as discussed in ref.~\cite{rev3,rev4}, we calculated the mean-field frequency ($\Omega$) and mean-ensemble frequency ($f$). The corresponding diagrams are illustrated in Figs.~\ref{mean}(a) and \ref{mean}(b) as a function of the coupling strength $(K)$ without and with additional perturbations $\varepsilon=0.0$ and $\varepsilon=0.3$. We observe the disordered values of the mean-field frequency in the dynchronized ($DS$) and partial synchronization ($PS$) states in the absence of perturbation ($\varepsilon=0.0$), as shown in Fig.~\ref{mean}(a) by the red linepoints. Additionally, the cluster oscillatory state or steady states show the presence of ordered coherent mean-field frequencies. We also noticed that on increasing the additional perturbation to $\varepsilon = 0.3$,  the disorderliness (see Fig.~\ref{mean}(b)) of the mean-field frequencies observed at the $DS$ and $PS$ states decreases. It is also important to notice that in both the cases of additional perturbations, the system(\ref{aeq2}) shows a lower range of coherent mean-ensemble frequencies. 
	
	
	\section{Robustness of observed dynamical states and their transitions in other dynamical models}
	\begin{figure*}[htb!]
		\hspace{-0.20cm}
		\includegraphics[width=16.0cm]{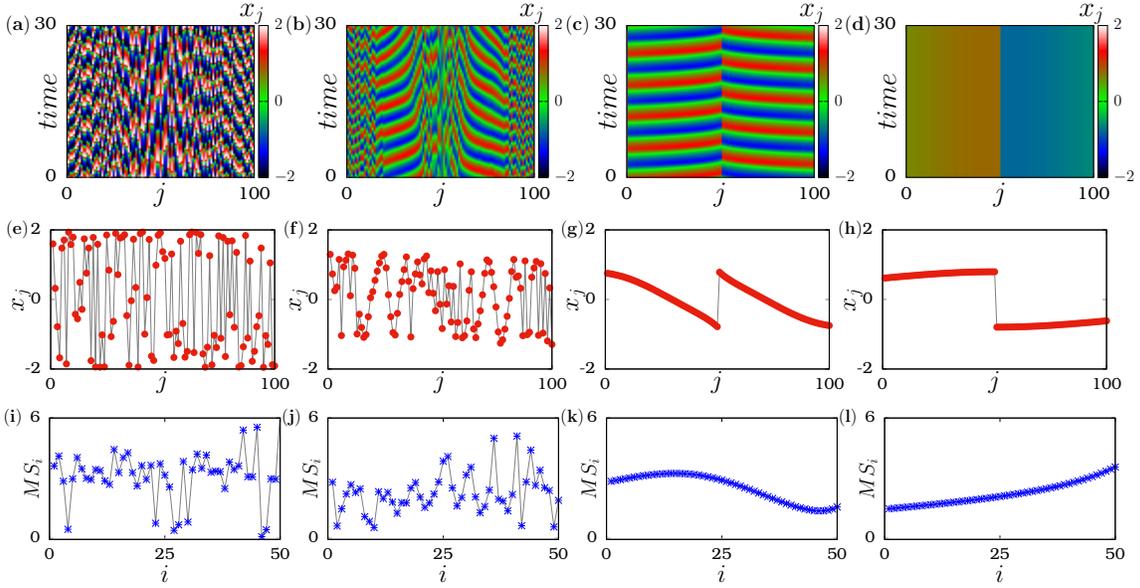}
		\vspace{-0.3cm}
		\caption { Spatio-temporal plots of globally coupled van der Pol oscillators in the absence of perturbation ($\varepsilon=0.0$) for (a) $K=0.05$ (desynchronized state), (b) $K=0.65$ (partial synchronization state), (c) $K=0.8$ (cluster oscillatory state), and (d) $K=1.0$ (oscillation death).  (e)-(h) are the snapshots of $x_j$ variables and (i)-(l) are the corresponding parity symmetry error $(MS_i)$. Other parameters: $b=2.0$, $\Delta=0.5$ and $N=100$.} 
		\label{vdpf1}
	\end{figure*}    
	\begin{figure*}[htb!]
		\hspace{-0.20cm}
		\includegraphics[width=16.0cm]{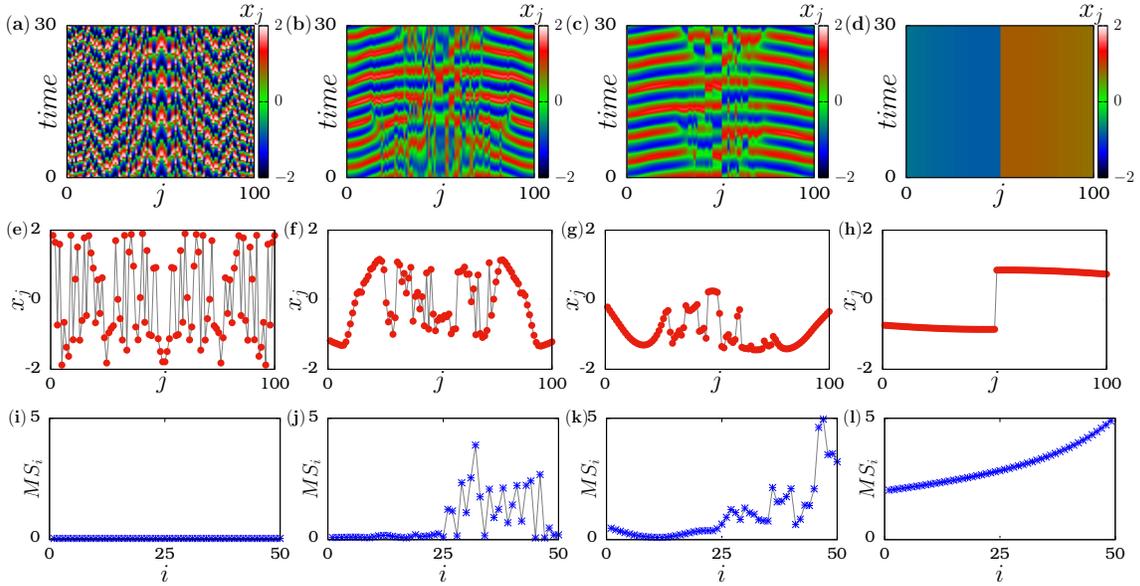}
		\vspace{-0.3cm}
		\caption {Spatio-temporal plots of globally coupled van der Pol oscillators in the presence of perturbation ($\varepsilon=0.3$) for (a) $K=0.05$ (incoherent mixed synchronization state), (b) $K=0.65$ (chimera state), (c) $K=0.8$ (chimera state), and (d) $K=1.0$ (oscillation death).  (e)-(h) are the snapshots of $x_j$ variables and (i)-(l) are the corresponding parity symmetry error $(MS_i)$. Other parameters are same as in Fig.~\ref{vdpf1}. } 
		\label{vdpf2}
	\end{figure*} 
	In order to confirm the generality of the observed dynamical states in other dynamical models, we choose the globally coupled van der Pol and R\"ossler nonidentical counter-rotating systems with additional perturbation. The corresponding dynamical transitions are detailed in the following subsections. 
	\subsection{Globally coupled van der Pol oscillators}

	\par The dynamical equations for the ring network of  globally coupled van der Pol oscillators are expressed as
	\begin{eqnarray}
	\dot{x_j} &=& \omega_j y_j  +\frac{K+\varepsilon}{N}\sum_{k=1}^{N}  ( {x_k}-\,  {x_j}), \nonumber \\
	\dot{y_i} & =& b(1-x_j^2)y_i+\frac{K-\varepsilon}{N}\sum_{k=1}^{N}  ( {y_k}-\,  {y_j}),
	\label{vdp}
	\end{eqnarray}
	where $b$ is the damping parameter which is chosen as $b=2$.  $\omega_j$ are the system frequencies that are distributed with uniform counter-rotating frequencies as detailed in Sec. II.  $K$ is the coupling strength and $\varepsilon$ is the additional perturbation.  The dynamical states observed from Eq.~(\ref{vdp}) are portrayed in Figs.~\ref{vdpf1} and \ref{vdpf2} in the absence and in the presence of additional perturbation.  Primarily, the dynamical states and their transitions are shown by fixing $b=2.0$, $\varepsilon=0.0$, and $\Delta=0.5$.  	

	At lower coupling strength  $K=0.05$, the existence of desynchronized state  is observed  as shown in Figs.~\ref{vdpf1}(a) and \ref{vdpf1}(e). On increasing the coupling strength to $K=0.65$, the oscillators in the system~(\ref{vdp}) get split into coherent and incoherent groups and exhibit a partial synchronization  state as shown in Figs.~\ref{vdpf1}(b) and \ref{vdpf1}(f). Here the incoherent behavior is observed at the edges and middle of the network. Further increment of $K$ to  $K=0.8$ results in the entire set of oscillators to form two independent clusters manifesting into a cluster oscillatory state.  In this state, the co-rotating oscillators form one cluster and the corresponding counter-rotating oscillators form the other cluster (see Figs.~\ref{vdpf1}(c) and \ref{vdpf1}(g)).  Furthermore,  at strong coupling strength  $K=2.0$, we find that the co- and counter-rotating oscillators populate into two different cluster oscillation death states as shown in  Figs.~\ref{vdpf1}(d) and \ref{vdpf1}(h).   
	
	\par	In addition, we illustrate the parity symmetry error of each dynamical state in Figs.~\ref{vdpf1}(i)-\ref{vdpf1}(l). By observing all the dynamical states in the absence of additional perturbation,  $\varepsilon =0.0$ (see Figs.\ref{vdpf1}(a) and \ref{vdpf1}(d)), we discover that each pair of oscillators yield a non-zero value of parity symmetry error indicating that the observed dynamical states including $DS,~PS,~COS$ and $COD$ states are of $P-$symmetry broken nature.  Subsequently, we perform a similar analysis in the presence of additional perturbation by fixing $\varepsilon=0.3$. As compared to the desynchronized state, here we find that each pair of the oscillators in the network preserves the parity symmetry (see Figs.~\ref{vdpf2}(a) and \ref{vdpf2}(e) for $K=0.05$) and results in the incoherent mixed synchronization state.  Therefore all the pairs of the oscillators show null value of $P-$symmetry error $MS_i$ in Fig.~\ref{vdpf2}(i). Further, the occurrence of chimera state is revealed while increasing $K$ to $K=0.65$ (see Figs.~\ref{vdpf2}(b) and \ref{vdpf2}(f)) and $K=0.8$ (see Figs.~\ref{vdpf2}(c) and \ref{vdpf2}(g)).  In the chimera state, some of the pairs of oscillators in the coherent and incoherent patches preserve the  $P-$symmetry ($MS_i=0$) while the others break the $P-$symmetry ($MS_i\ne0$) as illustrated in Figs.~\ref{vdpf2}(j) and \ref{vdpf2}(k). Furthermore, we discover similar kind of cluster oscillation death in  Figs.~\ref{vdpf2}(d), \ref{vdpf2}(h) and \ref{vdpf2}(l) at $K=2.0$ as observed in Figs.~\ref{vdpf1}(d), \ref{vdpf1}(h) and \ref{vdpf1}(l).
	As a result, our analysis clearly shows that the network of globally coupled van der Pol limit cycle oscillators also shows similar kind of dynamical states as that of the globally coupled Stuart-Landau oscillators.  
	
	\par  Furthermore, to provide additional support for our findings, we examine the dynamical behaviors using a chaotic system, specifically R\"ossler oscillators, in the following.   
	\subsection{\small \bf Globally coupled R\"ossler oscillators}
	In order to show the existence of partial synchronization and chimera states in a chaotic system, we consider the ring network of globally coupled R\"ossler oscillators which can be written as
	\begin{eqnarray}
	\dot{x_j} &=& -\omega_j y_j - z_j +\frac{K+\varepsilon}{N}\sum_{k=1}^{N}( {x_k}-\,  {x_j}), \nonumber \\
	\dot{y_j} & =& \omega_jx_j+a y_j +\frac{K-\varepsilon}{N}\sum_{k=1}^{N}( {y_k}-\,  {y_j}),  \nonumber \\ 
	\dot{z_j}& =& b+z_j(x_j-c)
	\end{eqnarray}
		\begin{figure}[h!] 
		\vspace{-0.0cm}
		\includegraphics[width=9.0cm]{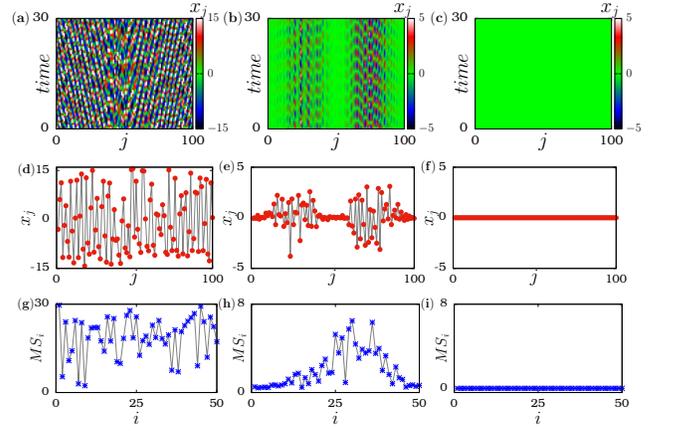}
		\vspace{-0.3cm}
		\caption {Spatio-temporal plots of globally coupled R\"ossler oscillators with non-identical counter-rotating frequencies in the absence of additional perturbation $(\varepsilon=0.0)$ for (a) $K=0.05$ (desynchronized state), (b) $K=0.1$ (partial synchronization state), (c) $K=0.15$ (amplitude death state),  (d)-(f) are the snapshots of $x_j$ variables, (g)-(i) are the corresponding parity symmetry error $(MS_i)$.  Other parameters: $a=0.2,~b=0.2,~c=10.0,~\Delta=0.1$ and $N=100$. } 
		\label{ros1}
	\end{figure} 
	\begin{figure}[h!] 
		\vspace{-0.0cm}
		\includegraphics[width=9.0cm]{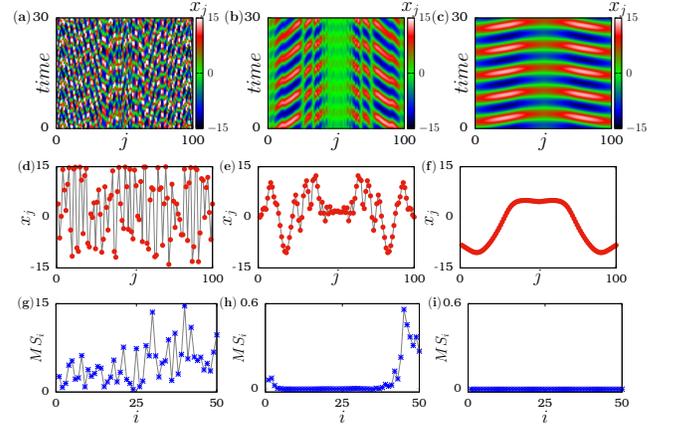}
		\vspace{-0.3cm}
		\caption {Spatio-temporal plots of globally coupled R\"ossler oscillators  with non-identical counter-rotating frequencies in the presence of additional perturbaton $(\varepsilon=0.05)$ for (a) $K=0.05$ (desynchronized state), (b) $K=0.1$ (chimera state), (c) $K=0.15$ (coherent mixed synchronization state),  (d)-(f) are the snapshots of $x_j$ variables, (g)-(i) are the corresponding parity symmetry error $(MS_i)$.  Other parameters are same as in Fig.~\ref{ros1}} 
		\label{ros2}
	\end{figure}
	where,  $a,~b$, and $c$ are the constant parameters and they are  chosen as $a=b=0.2$, and $c=10$.  $K$ is the coupling strength, $\varepsilon$ is the strength of additional perturbation, and $\omega_j$ are the system frequency.  The co- and counter-rotating frequencies are distributed with uniform distribution as detailed in Sec.~II.  Primarily, the dynamical behaviors are first analyzed for $\varepsilon=0.0$ and by fixing the coupling strength at different values.  For $K=0.01$, one can note   from Figs.~\ref{ros1}(a) and \ref{ros1}(d) that the oscillators are randomly distributed in the desynchronized state.  Then the existence of partial synchronization state with coexistence of coherent and incoherent behaviors is revealed from  Figs.~\ref{ros1}(b) and \ref{ros1}(e) for  $K=0.2$. Upon increasing the coupling strength, the system exhibits amplitude death state as seen in Figs.~\ref{ros1}(c) and \ref{ros1}(f). From the figures of  error of parity symmetry given in Figs.~ \ref{ros1}(g) and \ref{ros1}(h), we can note that the desynchronized state  and partial synchronization state are complete $P-$symmetry broken where all the oscillators take non-zero values of error. Due to trivial values of system variables, the  $P-$symmetry error of amplitude death state takes the null value as in Fig.~\ref{ros1}(i). 
	\par Furthermore, the effect of additional perturbation is explored by fixing $\varepsilon=0.05$ in Figs.~\ref{ros2}.  For $K=0.05$, we observed a desynchronized state as in Figs.~\ref{ros2}(a)-\ref{ros2}(d).  The corresponding $P-$symmetry error takes non-zero values of error which implies that the $P-$symmetry is broken in the $DS$ state, Fig.~\ref{ros2}(g). On increasing the coupling strength further, chimera states with the co-existence of coherent and incoherent behaviors are observed which is depicted in Figs.~\ref{ros2}(b) and \ref{ros2}(e). In such a state, some of the oscillators preserve the $P-$symmetry while the others do not preserve the $P-$symmetry.  As a result, the symmetry preserving pairs of oscillators in the network show zero error while the symmetry broken pairs show non-zero values of error as shown in Fig.~\ref{ros2}(h). On further increasing the coupling strength ($K=0.15$), we restore the oscillatory dynamics from amplitude death due to the influence of the additional perturbation. As a consequence, we observed a symmetry preserving cluster state that is a coherent mixed synchronization state (see Figs.~\ref{ros2}(c),\ref{ros2}(f), and \ref{ros2}(i).       
	
	\par Thus, from the above discussion, one can clearly observe that the results presented in the main text are more general and valid in all classes of limit cycle and chaotic systems.  Interestingly,  we identified that the additional perturbation leads to the onset the chimera behavior in nonidentical counter-rotating systems.   	
	
\end{document}